\documentstyle[epsfig]{lamuphys}
\makeatletter
\let\chapter\hid@chapter
\makeatother
\begin{document}
\pagenumbering{arabic}
\title{HST imaging of redshift $z>0.5$ 7C and 3C Quasars}

\author{Stephen Serjeant\inst{1}, Steve Rawlings\inst{2}, Mark Lacy\inst{2}}

\institute{$^1$Astrophysics Dept., Imperial College London, Blackett
Laboratory, 
Prince Consort Road, London SW7 2BZ, England\\
$^2$Astrophysics Dept., Oxford University, Nuclear and
Astrophysics Laboratory, 1 Keble Road, Oxford, OX1 3RH, England}

\maketitle

\begin{abstract}
We present preliminary results from HST imaging of 
radio-loud quasar hosts, covering a $\sim\times100$ 
range in radio luminosity but in a narrow redshift 
range ($0.5<z<0.65$).  The sample was selected from our
new, spectroscopically complete 7C survey and the 3CRR 
catalogue. Despite the very large radio luminosity range, 
the host luminosities are only weakly correlated (if at 
all) with radio power, perhaps reflecting a predominance 
of purely central engine processes in the formation of 
radio jets, and hence perhaps also in the radio-loud/-quiet
dichotomy at these redshifts. The results also contradict 
naive expectations from several quasar formation theories, 
but the host magnitudes support radio-loud Unified Schemes.

\end{abstract}
\section{Introduction}
The strong evolution of quasars from redshifts $z=2$ to $z=0$ is
very well documented but poorly understood. Their evolution
may reflect changing merger rates ({\it e.g.} Carlberg 1990)
or may reflect processes in galaxy formation {\it via} 
the changing formation efficiency of nuclear black holes ({\it e.g.}
Haehnelt \& Rees 1993). Alternatively, Small \& 
Blandford (1992) suggest that the largest galaxies, hosting the
brightest quasars, formed first (contrary to ``bottom-up''
heirarcichal structure formation); quasar evolution is then ascribed to a
luminosity-dependent transition from continuous to intermittent
activity. Physically, this could be interpreted an initial quasar
phase in newly-forming galaxies followed by merger-driven events. 
 
Each of these models has a wide parameter space in which to
accommodate strong positive quasar evolution, which reflects the
essential lack of a physical understanding. 
Nevertheless, the various classes of models 
make widely different predictions for the immediate environments of
the active nuclei. For example, in merger-driven quasar evolution
one expects tidal features
or disturbed morphologies in the host galaxies, possibly more
disturbed at higher $z$; in 
models where a quasar stage is common in bottom-up galaxy formation,
one expects that quasars of a given luminosity have increasingly
smaller host galaxies, with increasing redshift, but in the Small \&
Blandford (1992) model one predicts an opposite trend. 

Host galaxies also provide a useful test of radio-loud unified schemes
({\it e.g.} Antonucci 1993). If dust shrouded quasars are universal in
radiogalaxies, then there must be identical host galaxy properties in
radioquasars and radiogalaxies. However, in making such comparisons
one must ensure the quasar and
radiogalaxy samples are well-matched in some isotropic quantity, such
as hard X-ray luminosity or radio lobe luminosity. Unfortunately, the
only large, complete (published) sample of lobe-dominated quasars is
from the 3CR 
catalogue, in which radio luminosity is tightly correlated with
redshift. This makes it difficult to address the evolution of any
parameter, such as host galaxy luminosity, because of the possibility
of radio luminosity dependence. In
this paper we present a complete sample which 
breaks this degeneracy 
by comparing 3CR quasars with a new coeval, spectroscopically complete
sample $\sim100$ times fainter in radio luminosity. 


\begin{figure}
\vspace{-1.2cm}
\epsfig{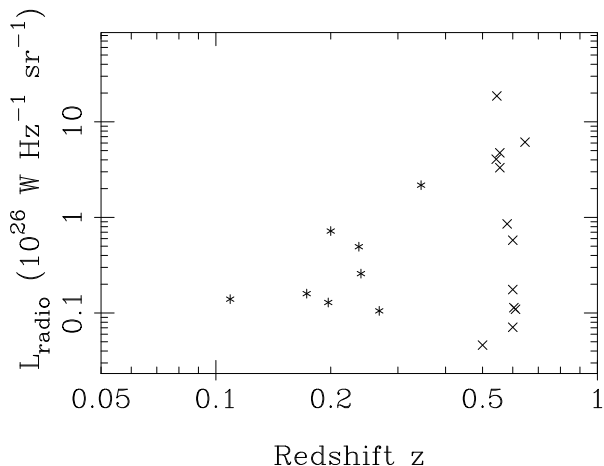}
\caption{Radio luminosity-redshift plane for our sample (crosses) and
steep-spectrum quasars from Dunlop {\it et al.} (1993)}
\end{figure}

\section{Data Aquisition and Analysis}
\subsection{Sample definition}
We have recently completed spectroscopic campaign on the INT and NOT
of steep-spectrum 
quasars (SSQs) from the $151$MHz 7C catalogue (McGilchrist {\it et al.}
1990), which has a limiting flux density of $S_{151}=0.1$Jy. 
This
sample is ideal for comparisons with coeval quasars from the $178$MHz
3CR catalogue (Laing, Riley \& Longair 1983), since the high flux
limit of $S_{178}=10$Jy in the latter gives a wide dispersion in radio
luminosity at any epoch. 
VLA snapshots of 7C obtained by us confirmed that the 3C and 7C SSQs
are from the same parent population; 
both samples consist of FRII sources with cores. 
As we will show below, this allows us to
decouple luminosity 
dependence from evolution. 

We were awarded twelve orbits to image a subsample of 3C and 7C SSQs
with the  
HST WFPC2. Our sample selection was driven by several competing
requirements: 
\begin{itemize}
\item spectroscopic completeness;
\item freedom (as far as possible) from beaming and gravitational
lensing biases; 
\item a narrow redshift interval, to counter any potential
differential evolution;
\item a choice of filter and redshift
range avoiding strong emission lines; 
\item as red a filter as possible to maximise the contrast between the
quasar and its host;
\item the highest accessible redshifts for reliable detection of host
galaxies. 
\end{itemize}

Our choice of SSQs (as opposed to flat-spectrum quasars)
ensured our samples were largely free from lensing and beaming since
the steep-spectrum radio fluxes are dominated by extended, optically
thin synchrotron emission. 

Based on crude models of PSF-subtracted frames, we estimated that SSQ
hosts should be detectable at $z\stackrel{<}{_\sim}0.6$ if Unified Scheme
predictions hold, {\it i.e.} that the host galaxies are giant
ellipticals. 
Our targets were therefore confined to the
interval $0.5<z<0.65$, to be imaged in the F675W filter which
successfully avoids the [O{\sc ii}] $3727$\AA, [O{\sc iii}] $5007$\AA\ 
and H$\beta$ emission lines. The complete
3CR and 7C samples contain 12 
steep-spectrum quasars in this redshift interval. The very wide range
in radio luminosity of our sample is 
demonstrated by the vertical dispersion in figure 1. 

Also plotted on
figure 1 are steep-spectrum 
quasars from the sample of Dunlop {\it et al.} (1993). Using samples
such as these we can
therefore make comparisons between well-matched quasars between epochs
({\it i.e.}, in a horizontal slice  of figure 1), which for the first
time are free of the luminosity dependence which marrs studies of 3C
alone. 

Moreover, even at the median redshift of our sample ($z\sim0.6$) the
quasar number density is 
already around an order of 
magnitude larger than the present ({\it e.g.} Dunlop \& Peacock
1990). Our sample 
is therefore (in principle at least) capable of probing any
environmental causes of quasar evolution. 

\begin{figure}
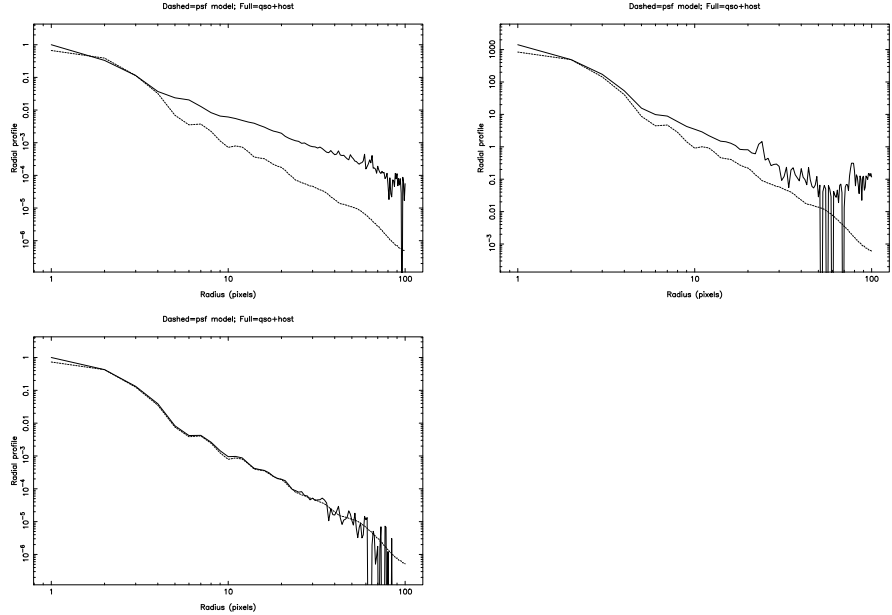

\epsfig{file=275.eps, angle=-90, width=6cm}
\epsfig{file=09.eps, angle=-90, width=6cm}
\epsfig{file=star.eps, angle=-90, width=6cm}
\caption{
Radial profiles for 3c275.1 (top left), the brightest host in our 
sample, and for 
a more typical target (7C2886, top right). The uprise at large radii in
the latter is due to a companion galaxy. In both cases the Tiny Tim
PSF is also plotted, normalised to the first unsaturated pixel. Also
shown is the empirical PSF compared to the Tiny Tim model (bottom). 
}
\vspace{-0.5cm}
\end{figure}

Fortuitously, two of the 3CR targets (3C334 and 3C275.1) also have
ground based detections 
of giant host galaxies, which was not realised when defining the
survey selection criteria. These detections provide important
corroborations of our photometry and quasar subtraction
technique. It is perhaps significant that these large host
galaxies, among the largest for any quasars, are at the highest
possible redshifts for optical ground-based detections. 

\subsection{Data aquisition}
For the majority of the targets we broke our observations of each
quasar (one orbit per quasar) into four separate exposures. Two of the
four exposures 
were offset by $5.5$ pixels 
in both directions, to assist with the positioning of the quasar in
the central pixel (recall that the Planetary Camera slightly
undersamples). Although this strategy might be expected to present
problems with the 
cosmic ray subtraction, we found that the cosmic rays could be
identified and removed adequately using the {\sc crrej} algorithm in the
{\sc IRAF imcombine} task, despite having only two frames per
position. 

Sky subtraction was achieved by selecting regions of the planetary
camera frame free of
bright objects, and constructing a histogram of pixel values (choosing
bin sizes and widths to minimise the digitisation effects); the sky
levels were then estimated by fitting Gaussians to these
histograms. 

\subsection{Point spread functions}
Since the Planetary Camera undersamples the point spread function, the
pixel to pixel variations 
may depend strongly on the position of the quasar within the central
pixel. We therefore expected that our QSO subtraction would be
improved by 
determinining this position {\it via} a
cross correlation of the quasar frame with a model oversampled point
spread function from Tiny Tim version 4.0 (Kirst 1995). The
oversampled model 
would need to be rebinned to the cruder Planetary Camera pixel scale
at each proposed centre, and a least squares solution found for the
central position. 
One small complication in this cross correlation is the existence of a
non-negligable pixel to pixel scattering function. The resampled point
spread functions must therefore be convolved with the appropriate kernel
from Kirst (1995) before comparisons are made with the data frame.

To complement the Tiny Tim models, we also constructed an empirical
point 
spread function from our standard star observations, masking out
saturated pixels before coaddition. An undersampled empirical point
spread function for this filter was unfortunately not available.

\begin{figure}
\vspace{-1.2cm}
\epsfig{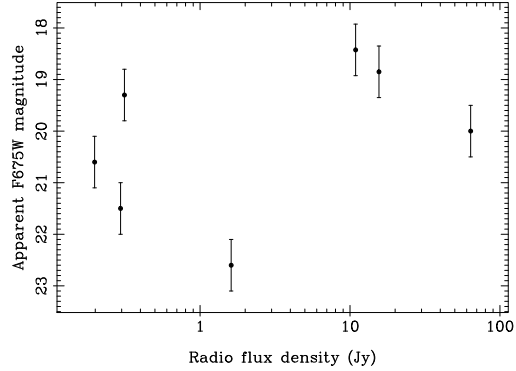}
\caption{Apparent F675W magnitudes of the hosts 
measured to date. Note the lack of, or only weak presence of, a
correlation with radio flux. The F675W magnitude system is about half
a magnitude fainter than Cousins R.}
\vspace{-0.5cm}
\end{figure}

\section{Results}

\subsection{Robustness of host magnitudes}
We attempted QSO
subtractions using PSFs obtained by three methods: an
arbitarily centred model PSF, the empirical PSF 
discussed above, and a PSF centred on the QSO by cross correlation
above with saturated pixels masked out. 
For the first two, we determined the PSF normalisation from the radial
profiles, using the first unsaturated pixel.
In the case of the
cross-correlation method, these normalisations were
in excellent agreement with those obtained independently from the
cross-correlation itself. 
Although specific features
such as diffraction spikes are better reproduced, the total
magnitudes were (reassuringly) found to be insensitive to the
PSF centering. 
We also attempted normalising to the
diffraction spikes, but 
the results were much less satisfactory. Nevertheless, the host
magnitudes were not found to be sensitive to the masking -or not- of
the diffraction spikes. 
The results for the quasars examined to date are 
shown in figures 2 and 3. The previous ground-based detections
of host galaxies in 3C275.1 and 3C334 are well reproduced by our
algorithms. From the dispersion between the different methods, we
estimate an error of $\sim\pm0.2$ in our host galaxy
magnitudes. 

We used azimuthal averaging to reach fainter flux
levels, as used by several other authors (figure 2). We might hope to
obtain a morphological classification 
from fits to the radial profiles ({\it e.g.} exponential disk, de
Vaucoleurs profile); 
this involves fitting to the curvature of the PSF-subtracted
profile, which is not as well constrained as the total host
magnitude. Furthermore, the reasonable 
inclusion of a bulge component to a disk model was found to introduce
too many free parameters. This will be discussed in more detail in
Serjeant {\it et al.} (1997), which also compares one- and
two-dimensional fits to our data.

\section{Discussion and Conclusions}
Remarkably, the host magnitudes are only weakly dependent -if at all-
on radio luminosity, despite the very wide dispersion in radio
luminosity ($\sim\times100$; see figure 3). This may reflect a
predominance of purely central 
engine processes in the formation of radio jets, and hence 
perhaps also in the radio-loud/-quiet dichotomy at these 
redshifts. If so, we predict that the radio-quiet and radio-loud hosts
should be similar by redshifts $z\simeq0.6$. 

We can compare figure 3 with the R magnitudes of radiogalaxies in our
redshift range from the samples of Lacy {\it et
al.} 1993 and Eales 1985. 
Both the radiogalaxies from these studies and our SSQ samples
were selected in an orientation-independent manner; 
these radiogalaxies have $18.5<R<21.0$ which is clearly
well reproduced by our data, supporting radio-loud
Unified Schemes.

The results also contradict naive expectations
from several quasar formation theories. In both the Haehnelt \& Rees (1994)
model (if applicable at this $z$) and the Small \& Blandford (1992)
model strong trends of host 
properties are expected with redshift (see introduction), whereas
giant ellipticals appear to host most radio-loud AGN at
$0.2<z<0.7$ ({\it e.g.} Taylor {\it et al.} 1996).

%

%
%
%

\end{document}